\documentclass{article}

\usepackage{arxiv}

\usepackage[utf8]{inputenc} 
\usepackage[T1]{fontenc}    
\usepackage{hyperref}       
\usepackage{url}            
\usepackage{booktabs}       
\usepackage{amsfonts}       
\usepackage{nicefrac}       
\usepackage{microtype}      
\usepackage{lipsum}		
\usepackage{natbib}
\usepackage{doi}
\usepackage{braket}
\usepackage{graphicx}
\usepackage{amsmath}
\usepackage{amssymb}
\usepackage{amsfonts}
\usepackage{braket}
\usepackage{float}
\usepackage[table,xcdraw]{xcolor}
\usepackage{cite}
\usepackage{subfig}

\usepackage{setspace}
\usepackage{multirow}

\title{Lackadaisical quantum walk in the hypercube to search for multiple marked vertices}

\author{ \href{https://orcid.org/0000-0002-6527-7065}{\includegraphics[scale=0.06]{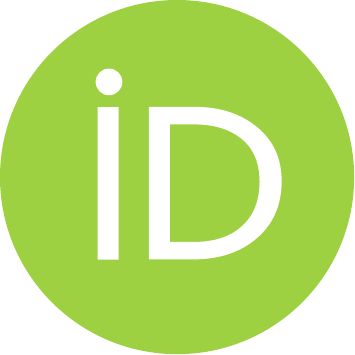}\hspace{1mm}Luciano S. de Souza}\thanks{95, R. Manuel de Medeiros, 35 - Dois Irmãos, Recife - PE} \\
	Departamento de Estat\'{i}stica e Inform\'{a}tica\\
	Universidade Federal Rural de Pernambuco\\
	Recife, Brasil \\
	\texttt{luciano.serafim@ufrpe.br} \\
	\And
	\href{https://orcid.org/0000-0002-2672-7801}{\includegraphics[scale=0.06]{orcid.pdf}\hspace{1mm}Jonathan H. A. de Carvalho} \\
	Centro de Inform\'{a}tica\\
	Universidade Federal de Pernambuco\\
	Recife, Brasil \\
	\texttt{jhac@cin.ufpe.br} \\
	\And
	\href{https://orcid.org/0000-0002-2131-9825}{\includegraphics[scale=0.06]{orcid.pdf}\hspace{1mm}Tiago A. E. Ferreira} \\
	Departamento de Estat\'{i}stica e Inform\'{a}tica\\
	Universidade Federal Rural de Pernambuco\\
	Recife, Brasil \\
	\texttt{tiago.espinola@ufrpe.br} \\
}

\renewcommand{\shorttitle}{LQW in the hypercube to search for multiple marked vertices}

\hypersetup{
pdftitle={A template for the arxiv style},
pdfsubject={q-bio.NC, q-bio.QM},
pdfauthor={David S.~Hippocampus, Elias D.~Striatum},
pdfkeywords={First keyword, Second keyword, More},
}

\begin{document}
\maketitle

\begin{abstract}

Adding self-loops at each vertex of a graph improves the performance of quantum walks algorithms over loopless algorithms. Many works approach quantum walks to search for a single marked vertex. In this article, we experimentally address several problems related to quantum walk in the hypercube with self-loops to search for multiple marked vertices. We first investigate the quantum walk in the loopless hypercube. We saw that neighbor vertices are also amplified and that approximately $1/2$ of the system energy is concentrated in them. We show that the optimal value of $l$ for a single marked vertex is not optimal for multiple marked vertices. We define a new value of $l = (n/N)\cdot k$ to search multiple marked vertices. Next, we use this new value of $l$ found to analyze the search for multiple marked vertices non-adjacent and show that the probability of success is close to $1$. We also use the new value of $l$ found to analyze the search for several marked vertices that are adjacent and show that the probability of success is directly proportional to the density of marked vertices in the neighborhood. We also show that, in the case where neighbors are marked, if there is at least one non-adjacent marked vertex, the probability of success increases to close to $1$. The results found show that the self-loop value for the quantum walk in the hypercube to search for several marked vertices is $l = (n / N) \cdot k $.

\end{abstract}
\section{Introduction}
\label{sec:introduction}

According to \citet{shenvi2003quantum}, quantum walks provide one of the most promising features, an intuitive framework for building new quantum algorithms. They were pioneers in designing a quantum search algorithm on the hypercube based on quantum random walks \citep{potovcek2009optimized}. Recent works have used the quantum walks to search weights and train artificial neural networks \citep{de2019quantum,desouza2021classical}.

The topology of the structure where the walk is applied considerably affects the evolution of the walker \citep{wang2017adjustable}. Therefore, many works are developed to improve the performance of quantum walks, quantum search algorithms in different structures: one-dimensional, two-dimensional, and multidimensional grids, complete and bipartite graphs, among others \citep{bezerra2021quantum,de2020impacts,nahimovs2019adjacent,rhodes2019quantum}.

Quantum walk modification proposals are also made to improve their performance. For example, \citet{wong2018faster} added to each vertex of a two-dimensional grid a self-loop, so the walker has some probability of staying put, achieving an improvement over the algorithm without self-loop \citep{ambainis2004coins}. 

\citet{rhodes2020search} proposed an ideal weight for all vertex-transitive graphs with a single marked vertex such that the ideal self-loop weight is equal to the degree of the loopless graph divided by the total number of vertices. \citet{potovcek2009optimized} observed that the nearest neighbors are also presented with high probability and \citet{nahimovs2019lackadaisical} that adjacent vertices can be hard to find by quantum walks.

In this way, we investigate whether the optimal value of $l = (d / N)$ for a single marked vertex is optimal for multiple marked vertices, where $d$ is the degree of the loopless vertex and $N$ is the number of vertices. We analyzed the quantum walk on hypercube without self-loop and with self-loop. We analyzed the quantum walk on the hypercube for multiple marked adjacent and non-adjacent vertices. Finally, we find an optimal value of $l$ for a quantum walk in the hypercube with multiple marked vertices.

This paper is organized as follows. In Section \ref{sec:quantum-walk}, we present some concepts about quantum walks and specifically the quantum walk on the hypercube. In Section \ref{sec:analyzing-the-quantum-walk-on-the-hypercube}, we characterize the probability distribution along with the space, adjust the self-loop weight for multiple marked vertices, and search for adjacent marked vertices. Finally, in Section \ref{sec:conclusions} is the conclusion.
\section{Quantum Walk}
\label{sec:quantum-walk}

The processing of quantum information is governed by quantum mechanics or quantum physics \citep{singh2016evolution}. Quantum computing study the processing of this information \citep{nielsen2002quantum,yanofsky2008quantum,mcmahon2007quantum}. Quantum walks are the quantum counterpart of classical random walks. Discrete and continuous-time quantum walks are the advanced tools used to build quantum algorithms \citep{aharonov1993quantum, ambainis2012search}. The main feature that differentiates these two types of quantum walks is the timing used to applying the evolution operators. In the quantum walk in continuous time, the evolution operator is applied at any time, whereas the quantum walks in discrete time, the evolution operator is applied in discrete time steps \citep{venegas2012quantum}. The quantum walk evolution in the discrete-time process occurs by the successive applications of a unitary evolution operator $U$ that acts on the Hilbert space 

\[\mathcal{H} = \mathcal{H}^{C} \otimes \mathcal{H}^{S}.\] 

The coin space $\mathcal{H}^{C}$ is the Hilbert space associated with a quantum coin, and the walker's space $\mathcal{H}^{S}$ is the Hilbert space associated with the position of the nodes in a graph, for example. The evolution operator $U$ is defined in Equation \ref{eq:evolution-operator}.

\begin{equation}
\label{eq:evolution-operator}
    U = S(C\otimes I_{N})
\end{equation}
where, $S$ is the shift operator, i.e., a permutation matrix that acts in the walker's space based on the state of the coin space. The unitary matrix $C$ is the coin operator \citep{shenvi2003quantum}.
Therefore, the equation of evolution represented by a quantum walk at time $t$ is given by 

\[\ket{\Psi(t)} = U^{t}\ket{\Psi(0)}\].

\subsection{Quantum walk on the hypercube}
\label{sec:quantum-walks-hypercube}

According to \citet{venegas2012quantum}, the hypercube is defined as an undirected graph of degree $n$ and $N = 2^{n}$ nodes. Each node is represented by an $n$-bit binary string. Two nodes $\Vec{x}$ and $\Vec{y}$ are connected by an edge if the Hamming distance between them is $1$, i.e., $\left| \Vec{x} - \Vec{y} \right| = 1$. This means that $\vec{x}$ and $\vec{y}$ only differ in a single bit. The expression $\left | \vec{x} \right |$ is the Hamming weight of $\vec{x}$. The Hilbert space associated with the quantum walk on the hypercube is

\[\mathcal{H} = \mathcal{H}^{n} \otimes \mathcal{H}^{2^{n}},\]
where $\mathcal{H}^{n}$ is the Hilbert space associated with the quantum coin space, and $\mathcal{H}^{2^{n}}$ is the Hilbert space associated with nodes on the hypercube.

According to \citet{shenvi2003quantum}, in a $d$-dimensional hypercube, the $d$ directions specify the coin state. \citet{kempe2002quantum} defines that directions can be labeled by the $n$ base-vectors $\{\ket{0}, \ket{1}, \dots, \ket{n-1}\}$ on the hypercube which corresponding to the $n$ vectors of Hamming weight $1$. These $n$ vectors are represented by the states $\{\ket{e_{0}},\ket{e_{1}},\dots,\ket{e_{n-1}}\}$, where $e_{d}$ has a $1$ in the $d$-th bit. The shift operator $S$ described in Equation \ref{eq:shift-operator-hypercube} acts mapping a state $\ket{d,\vec{x}} \rightarrow \ket{d,\vec{x}\oplus \vec{e_{d}}}$.

\begin{equation}
    \label{eq:shift-operator-hypercube}
    S = \sum_{d=0}^{n-1}\sum_{\vec{x}} \ket{d,\vec{x}\oplus \vec{e_{d}}}\bra{d,\vec{x}}
\end{equation}

The initial state of the quantum walk in the hypercube is defined according to Equation \ref{eq:initial-state-quantum-walk-hypercube} as an equal superposition over all $N$ nodes and $n$ directions.

\begin{equation}
\label{eq:initial-state-quantum-walk-hypercube}
    \ket{\Psi(0)} = \frac{1}{\sqrt{n}} \sum_{d=0}^{n-1}\ket{d}\otimes \frac{1}{\sqrt{N}} \sum_{\vec{x}} \ket{\vec{x}}
\end{equation}

According to \citet{rhodes2020search}, the hypercube was the first graph in which quantum walks were researched. In their work, \citet{shenvi2003quantum} presented a quantum search algorithm based on the random walk quantum architecture. In this article, we are based on the approach used by \citet{wong2018faster}. The pure quantum walk (without search) evolves by repeated applications from the evolution operator described in Equation \ref{eq:evolution-operator}, where $C$ is Grover's ``diffusion'' operator on the coin space and is given by

\begin{equation}
\label{eq:grovers-coin}
    C = 2\ket{s^{C}}\bra{s^{C}} - I_{n}
\end{equation}
where, $I_{n}$ is the identity operator, $n$ is the vertex degree loopless, and $\ket{s^{C}}$ is the equal superposition over all $n$ directions \citep{moore2002quantum,shenvi2003quantum}, i.e.,

\begin{equation}
    \ket{s^{C}} = \frac{1}{\sqrt{n}} \sum_{d=0}^{n-1} \ket{d}.
\end{equation}
We include a query to the ``Grover oracle'', described in Equation \ref{eq:query-oracle-grover}, at each step of the quantum walk.

\begin{equation}
\label{eq:query-oracle-grover}
    U' = U \cdot (I_{n} \otimes Q)
\end{equation}
where, $Q = I_{N} - 2\ket{\omega}\bra{\omega}$, and $\ket{\omega}$ means the marked vertex. The system is initiated according to the initial state presented in Equation \ref{eq:initial-state-quantum-walk-hypercube}.
\section{Analyzing the quantum walk on the hypercube}
\label{sec:analyzing-the-quantum-walk-on-the-hypercube}

In this section, we experimentally analyze the quantum walk on the hypercube searching for multiple marked vertices. The simulations and the obtained results are detailed in the following subsections.

\subsection{Characterizing the probability distribution along the space}
\label{sec:loopless-version}

Previous works showed there is an amplification in the solution neighborhood, which interferes with the amplification of the solutions by the quantum walk on the hypercube \citep{shenvi2003quantum,potovcek2009optimized,nahimovs2019adjacent}. Initially, it is necessary to understand how the probability amplitudes are distributed in the search space and how the quantum walk evolves in the hypercube over time considering the impacts caused by the solution neighborhood.

\begin{figure}
\centerline{\includegraphics[width=0.7\textwidth]{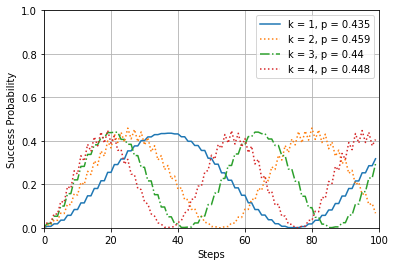}}
\caption{Success probability after 100 steps in a hypercube with $1024$ nodes. The solid blue curve is the success probability for one solution. The dotted orange curve is the success probability for two solutions. The dot-dashed green curve is the success probability for three solutions. The dotted red curve is the success probability for four solutions.} \label{fig:success-probability-one-two-three-solutions}
\end{figure}

Figure \ref{fig:success-probability-one-two-three-solutions} shows the probability of success after one hundred steps for the quantum walk in the hypercube with one, two, three, and four arbitrarily marked vertices. Although the search algorithm is able to amplify the probability amplitudes of the marked vertices, if a measurement is performed, the probability of finding one of the solutions is still unsatisfactory. Another interesting aspect that can be observed is that as the number of marked vertices increases, the speed of amplification the probability amplitudes also increases. However, it is necessary to increase the probability amplitudes of the marked vertices.

\begin{figure}
\centering
\subfloat[]{\includegraphics[width = 3.2in]{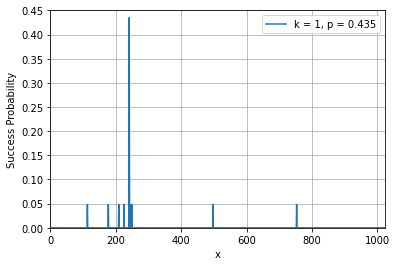}
\label{fig:probability-distribution-a}}\vspace{-5pt}
\subfloat[]{\includegraphics[width = 3.2in]{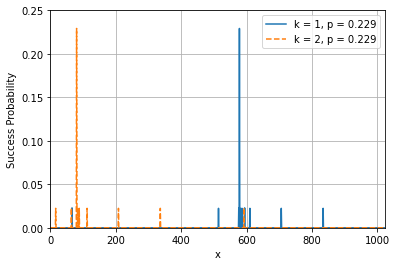}\label{fig:probability-distribution-b}}\\
\subfloat[]{\includegraphics[width = 3.2in]{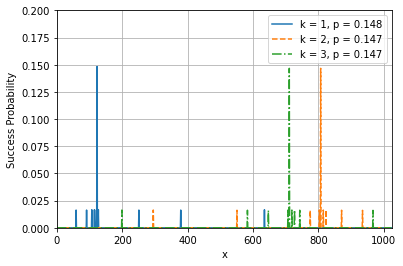}\label{fig:probability-distribution-c}}
\subfloat[]{\includegraphics[width =3.2in]{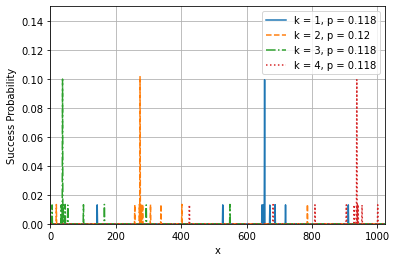}\label{fig:probability-distribution-d}}
\caption{Probability distribution of the  quantum walk after the number of iterations necessary to reach the maximum value of the probability amplitude with $n = 10$ and $N = 1024$ vertices. The y-axis values are at different ranges to improve visualization. \protect\subref{fig:probability-distribution-a} solid blue bar show the probability distribution for one marked vertex. \protect\subref{fig:probability-distribution-b} solid blue bar and orange dashed bar show the probability distribution for two marked vertices. \protect\subref{fig:probability-distribution-c} solid blue bar, orange dashed bar and green dash-dot bar show the probability distribution for three marked vertices. \protect\subref{fig:probability-distribution-d} solid blue bar, orange dashed bar, green dash-dot bar and dotted red bar show the probability distribution for four marked vertices.}
\label{fig:probability-distribution-without-selfloop}
\end{figure}

Figure \ref{fig:probability-distribution-without-selfloop} shows the probability distributions of the marked vertices only after the number of iterations necessary to reach the maximum value of the probability amplitude close to $1/2$. As \citet{potovcek2009optimized} noted in their work, we also note that the set of neighbors have a high probability. If we add the amplitudes of the neighbor's vertices, the values are compatible with the amplitude value of the marked vertex. We conclude that a considerable part of the energy, approximately $1/2$, is retained in the neighbors of the marked vertices. Figure \ref{fig:probability-distribution-d}, shows the probability distribution of four marked vertices. Note that the amplitudes of each vertex have their maximum and a neighborhood region. The x-axis distribution is the relative position of the position on the hypercube. It explains why even increasing the number of marked vertices, the success probabilities do not reach values above $1/2$.

\begin{figure}
\centering
\subfloat[]{\includegraphics[width = 3.2in]{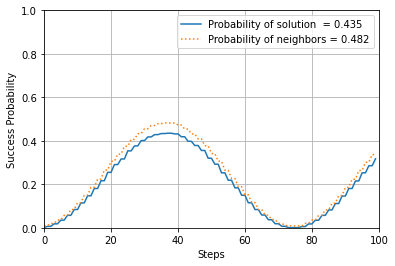}\label{fig:success-probability-measurement-a}} 
\subfloat[]{\includegraphics[width = 3.2in]{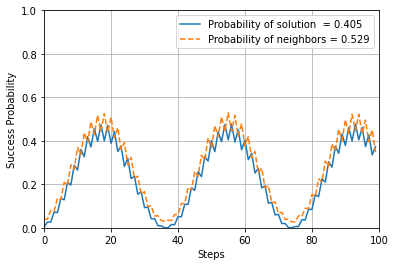}\label{fig:success-probability-measurement-b}}
\caption{Probability of success after 100 steps with $n = 10$ and $N = 1024$ vertices. \protect\subref{fig:success-probability-measurement-a} shows the probability of success for one marked vertex and its neighbors. \protect\subref{fig:success-probability-measurement-b} shows the probability of success for four marked vertices and their neighbors.}
\label{fig:probability-distribution-without-selfloop-measurement}
\end{figure}

Figure \ref{fig:probability-distribution-without-selfloop-measurement} shows the success probability for the quantum walk with one and four marked vertices after one hundred steps. Figure \ref{fig:success-probability-measurement-a} shows the behavior of the success probability of one marked vertex, the solid blue curve, and its neighbors, which is the dotted orange curve. If a measurement is performed, the probability of getting a neighbor vertex is greater than getting a marked vertex. With probability above $90\%$, you get the solution or a vertex that is one step away from the solution. Figure \ref{fig:success-probability-measurement-b} shows the behavior of the success probability of four marked vertices, the solid blue curve, and their neighbors, the dotted orange curve. Note that in a step when the probability of success of the marked vertices is high, the probability of success of the neighbors decreases, and in the next step, when the probability of success of the neighbors is high, the probability of success of the marked vertices decreases. Because of this behavior, if a measurement is performed, the probability of getting a neighbor is high. This happens in Figure \ref{fig:success-probability-measurement-a} but more smoothly.

Observing these results, we must consider the probability $p$ of obtaining a marked vertex and the probability $p' = (1 - p)$ of obtaining an unmarked vertex which is the sum of the probabilities of the $(N - k)$ vertices, where $k$ is the number of marked vertices. These results are shown in Table \ref{tab:probabilities-of-success-of-marked-and-unmarked-vertices}. Note the column of the value of $p'$, which is composed of the value of the amplitudes of the neighbors and the amplitude of the vertices that are neither neighbors nor marked. The probability of the walker finding a region is high because the energy is concentrated in the neighboring region. It is concluded that the amplification of the neighborhood around the marked vertices interferes with the probability of success of finding a target vertex.

\begin{table}
\setlength{\tabcolsep}{30pt}
\centering
\caption{Probabilities of success of marked and unmarked vertices.}
\label{tab:probabilities-of-success-of-marked-and-unmarked-vertices}
\begin{tabular}{cccc}
\hline
\multicolumn{4}{c}{Probabilities of success} \\ \hline
\multicolumn{1}{c|}{\multirow{2}{*}{Figure}}               & \multicolumn{1}{c|}{\multirow{2}{*}{$p$}} & \multicolumn{2}{c}{$p' = (1 - p)$}       \\ \cline{3-4} 
\multicolumn{1}{c|}{}                                      & \multicolumn{1}{c|}{}                     & \multicolumn{1}{c|}{Neighbors} & Neither \\ \hline
\multicolumn{1}{c|}{\ref{fig:probability-distribution-a}}        & \multicolumn{1}{c|}{$43.5\%$}             & \multicolumn{1}{c|}{$48.2\%$}  & $8.3\%$   \\
\multicolumn{1}{c|}{\ref{fig:probability-distribution-b}}        & \multicolumn{1}{c|}{$45.8\%$}            & \multicolumn{1}{c|}{$45.5\%$} & $8.7\%$   \\
\multicolumn{1}{c|}{\ref{fig:probability-distribution-c}}        & \multicolumn{1}{c|}{$44.2\%$}             & \multicolumn{1}{c|}{$48.4\%$} & $7.4\%$   \\
\multicolumn{1}{c|}{\ref{fig:probability-distribution-d}}        & \multicolumn{1}{c|}{$47.4\%$}             & \multicolumn{1}{c|}{$44.5\%$} & $8.1\%$   \\
\multicolumn{1}{c|}{\ref{fig:success-probability-measurement-a}} & \multicolumn{1}{c|}{$43.5\%$}             & \multicolumn{1}{c|}{$48.2\%$} & $8.3\%$   \\
\multicolumn{1}{c|}{\ref{fig:success-probability-measurement-b}} & \multicolumn{1}{c|}{$40.5\%$}             & \multicolumn{1}{c|}{$52.9\%$}  & $6.6\%$   \\ \hline
\end{tabular}
\end{table}

\subsection{Adjusting the self-loop weight for multiple marked vertices}
\label{sec:adjusting-the self-loop-weight-for multiple-marked-vertices}

Many works have been proposed with the purpose of improving the search capacity of quantum algorithms. According to \citet{wong2015grover}, adding a self-loop to each vertex boosts the success probability from $1/2$ to $1$. A modification to the initial state in the Equation \ref{eq:initial-state-quantum-walk-hypercube} and to Grover's coin in the Equation \ref{eq:grovers-coin} is needed so that the self-loop can be added. The addition of the self-loop is described in Equation \ref{eq:add-self-loop}. Thus, the coin space is now an $(n + 1)$-dimensional space \citep{rhodes2020search}.

\begin{equation}
\label{eq:add-self-loop}
    \ket{s^{C}} = \frac{1}{\sqrt{n + l}} \left ( \sqrt{l}\ket{\circlearrowleft} + \sum_{d=0}^{n-1}\ket{d} \right )
\end{equation}

One of the concerns when adding a self-loop at each vertex is knowing the best self-loop value. More specifically, in the case of the quantum walk on the hypercube, \citet{rhodes2020search} proposed an optimal self-loop value

\begin{equation}
\label{eq:optimal-self-loop}
    l = \frac{d}{N},
\end{equation} 
where $d$ is equal to the degree of the loopless graph and $N$ is the number of vertices in the hypercube. Recently, two works showed that inserting the number of marked vertices in  calculating the self-loop value optimizes quantum walks. \citet{decarvalho2021applying} shows that the optimal value of the self-loop for quantum walks in $D$-dimensional grids with multiple marked vertices is 

\[l = \frac{2Dm}{N},\]
where $2D$ is the number of movements the walker can do, not counting the self-loop, $m$ the number of marked vertices, and $N$ the number of vertices of the grid. \citet{nahimovs2021lackadaisical} shows that for different types of two-dimensional grids - triangular, rectangular, and honeycomb the optimal self-loop value is also,

\[ l = \frac{m\cdot d}{N} \]
where $d$ is the degree of the vertex, $m$ is the number of marked vertices, and $N$ is the number of vertices of the grid.

\begin{figure}
\centering
\subfloat[]{\includegraphics[width = 3.2in]{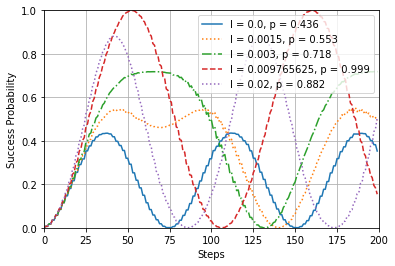}\label{fig:success-probability-a}} 
\subfloat[]{\includegraphics[width = 3.2in]{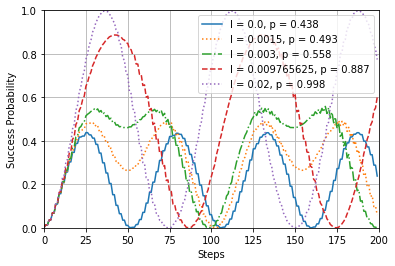}\label{fig:success-probability-b}}\\
\subfloat[]{\includegraphics[width = 3.2in]{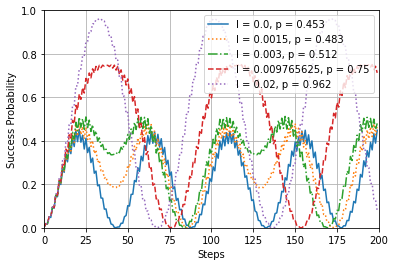}\label{fig:success-probability-c}}
\subfloat[]{\includegraphics[width = 3.2in]{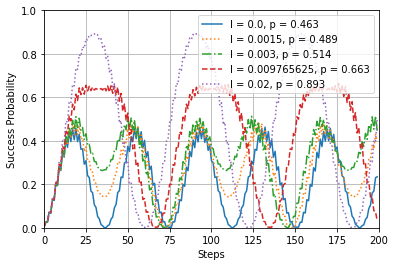}\label{fig:success-probability-d}}
\caption{Comparison between multiple self-loops values and $l = (n/N)$. \protect\subref{fig:success-probability-a} shows the success probability for one marked vertex. \protect\subref{fig:success-probability-b} shows the success probability for two marked vertices. \protect\subref{fig:success-probability-c} shows the success probability for three marked vertices. \protect\subref{fig:success-probability-d} shows the success probability for four marked vertices.}
\label{fig:success-probabilitya-b-c-d}
\end{figure}

Figure \ref{fig:success-probability-a} shows the probability of success after two hundred steps for one marked vertex. Here, the values of $l$ were the same as used by Rhodes. The dashed red curve has the optimum value of $l$. Our interest was to investigate whether the value of $l$ described in Equation \ref{eq:optimal-self-loop} also improved the walk results for a number ($k>1$) of marked vertices. For this, we performed three more experiments where we increased the number of marked vertices up to four. As we added the marked vertices the success probability of the dashed red curve decreased to $88.7\%$ (\ref{fig:success-probability-b}) while the success probability of the dotted purple curve increased to $99.8\%$ (\ref{fig:success-probability-b}) but then also decreased to $96.2\%$ (\ref{fig:success-probability-c}) and $89.3\%$ (\ref{fig:success-probability-d}). It indicates that a new value of $l$ is required when the number of marked vertices increases. To find the optimal self-loop for multiple marked vertices, we defined a set of values in the form $l' = (\alpha \cdot l)$, where $\alpha \in \mathbb{N}$.

\begin{figure}
\centerline{\includegraphics[width=0.7\textwidth]{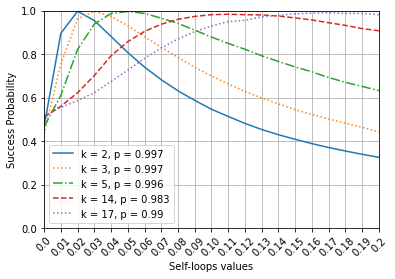}}
\caption{Investigation to set the value of $l$ for multiple marked vertices.} \label{fig:investigation-value-selflopp}
\end{figure}

Figure \ref{fig:investigation-value-selflopp} compares the probability of success for a set of marked vertices, $k = \{2, 3, 5, 14, 17\}$, these vertices were chosen randomly as well as their number. The self-loop values for these vertex numbers are $\alpha \cdot l$, where $l = (d/N)$ and $\alpha = \{1, 2, 3,...\}$. Note that the curves have their maximum points exactly at the locations on the x-axis where the $l'$ values are. We can conclude that the value of ($\alpha = k$). Therefore, we can set the value of $l$ for multiple marked vertices for the quantum walk in the hypercube, 

\begin{equation}
    l' = \frac{n}{N} \cdot k
\end{equation}
where $n$ is equal to the degree of the loopless vertex of the hypercube, $N$ the number of vertices in the hypercube, and $k$ the number of marked vertices. The self-loop value shown by \citet{nahimovs2021lackadaisical} for the quantum search in various types of two-dimensional grids coincides with the optimal self-loop value for the search for a quantum walk in the hypercube.

Figure \ref{fig:investigation-value-selflopp} shows that, as the values of $l$ approach the optimal value, the probability of success of the curve also approaches its maximum value. We can observe this behavior in Table \ref{tab:success-probability-for-one-to-ten-marked-vertices} which shows the probability of success for multiple values of $l$ and multiple marked vertices. Consider the values of the main diagonal, which are the maximum success probabilities for each $l = (n/N)\cdot k$.

\begin{table}
\setlength{\tabcolsep}{7pt}
\footnotesize
\centering
\caption{Probability of success for multiple values of $l$.}
\begin{tabular}{l|llllllllll}
\hline
\multirow{2}{*}{$l = (n/N)\cdot k$} & \multicolumn{10}{c}{Number of marked vertices} \\ \cline{2-11}

& \multicolumn{1}{c}{1} & \multicolumn{1}{c}{2} & \multicolumn{1}{c}{3} & \multicolumn{1}{c}{4} & \multicolumn{1}{c}{5} & \multicolumn{1}{c}{6} & \multicolumn{1}{c}{7} & \multicolumn{1}{c}{8} & \multicolumn{1}{c}{9} & \multicolumn{1}{c}{10} \\ \hline
\multicolumn{1}{l|}{(n/N)*1} & \textbf{0.999} & 0.888 & 0.75 & 0.663 & 0.775 & 0.592 & 0.575 & 0.576 & 0.589 & 0.55 \\
\multicolumn{1}{l|}{(n/N)*2} & 0.888 & \textbf{0.998} & 0.958 & 0.886 & 0.815 & 0.9 & 0.705 & 0.672 & 0.639 & 0.624 \\
\multicolumn{1}{l|}{(n/N)*3} & 0.749 & 0.959 & \textbf{0.998} & 0.976 & 0.934 & 0.886 & 0.941 & 0.792 & 0.857 & 0.727 \\
\multicolumn{1}{l|}{(n/N)*4} & 0.64 & 0.888 & 0.978 & \textbf{0.998} & 0.975 & 0.954 & 0.922 & 0.885 & 0.847 & 0.813 \\
\multicolumn{1}{l|}{(n/N)*5} & 0.555 & 0.816 & 0.937 & 0.986 & \textbf{0.996} & 0.989 & 0.966 & 0.943 & 0.912 & 0.883 \\
\multicolumn{1}{l|}{(n/N)*6} & 0.49 & 0.75 & 0.888 & 0.958 & 0.989 & \textbf{0.996} & 0.991 & 0.973 & 0.953 & 0.928 \\
\multicolumn{1}{l|}{(n/N)*7} & 0.438 & 0.691 & 0.84 & 0.926 & 0.969 & 0.992 & \textbf{0.996} & 0.99 & 0.978 & 0.983 \\
\multicolumn{1}{l|}{(n/N)*8} & 0.395 & 0.641 & 0.794 & 0.888 & 0.944 & 0.895 & 0.991 & \textbf{0.993} & 0.99 & 0.988 \\
\multicolumn{1}{l|}{(n/N)*9} & 0.361 & 0.596 & 0.75 & 0.852 & 0.915 & 0.957 & 0.978 & 0.993 & \textbf{0.994} & 0.982 \\
\multicolumn{1}{l|}{(n/N)*10} & 0.331 & 0.554 & 0.711 & 0.816 & 0.888 & 0.935 & 0.966 & 0.982 & 0.99 & \textbf{0.996} \\ \hline
\end{tabular}
\label{tab:success-probability-for-one-to-ten-marked-vertices}
\end{table}

\begin{figure}
\centerline{\includegraphics[width=0.7\textwidth]{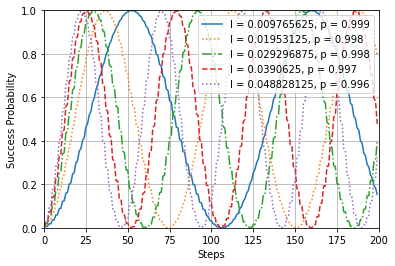}}
\caption{Probability of success after 200 steps. Solid blue curve, $k = 1$. Dotted orange curve, $k = 2$. Green dash-dot curve, $k = 3$. Red dashed curve, $k = 4$. Dotted purple curve, $k = 5$ } \label{fig:success-probability-five_solutions-with-optimal-selfloop}
\end{figure}

Table \ref{tab:success-probability-for-one-to-ten-marked-vertices} shows the relationship between the self-loop value and the number of marked vertices. We observe the relationship between the self-loop value and the number of marked vertices. Note that when the values of $l$ approach the optimal values for each number of marked vertices, there is an improvement in the probability amplitude. Figure \ref{fig:success-probability-five_solutions-with-optimal-selfloop} shows the probability of success after two hundred steps for multiple marked vertices. We can conclude that for cases where there is more than one marked vertex, the optimal value of $l = (n/N)\cdot k$.

\subsection{Searching for adjacent marked vertices}
\label{sec:searching-for-adjacent-marked-vertices}

The results found in the previous sections refer to the search for non-adjacent marked vertices, i.e., $\left |\vec{\omega_{i}} - \vec{\omega_{j}}\right | \neq 1$ the Hamming distance from vertex $\vec{\omega_{i}}$ and all other marked vertices is different from $1$. \citet{nahimovs2019adjacent} shows in their work that for quantum walks in the hypercube if the search space contains marked neighbors vertices, the search can be drastically affected. The authors considered two sets, one with two adjacent marked vertices and the other with two non-adjacent marked vertices. In the first case, the two adjacent marked vertices are $M = \{0, 1\}$. The absolute value of the overlap remained close to $1$, and the probability remains close to the initial state probability. In the second case, the two non-adjacent marked vertices are $M = \{0, 3\}$. The behavior on this one is different, the same behavior as the solid blue curve in Figure \ref{fig:success-probability-measurement-a}.

As the addition of self-loop in the quantum walk in the hypercube improved the search for multiple non-adjacent marked vertices, we investigated the case where the marked vertices are adjacent. We consider ten sets of vertices, $M = [\{0,1\},\{0,1,2\},\cdots,\{0,1,2,4,8,\cdots,256,512\}]$, i.e., all vertices adjacent to the vertex $0$. We add one more vertex to the set of marked vertices on each new walk until the number of vertices in $M$ is equal to the degree $n$ of the vertex.

\begin{figure}
\centering
\subfloat[]{\includegraphics[width = 3.2in]{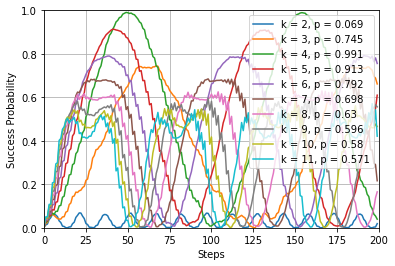}\label{fig:success-probability-marked-neighbors-l-N-l-N-k-a}} 
\subfloat[]{\includegraphics[width = 3.2in]{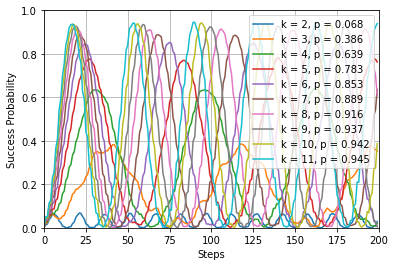}\label{fig:success-probability-marked-neighbors-l-N-l-N-k-b}}
\caption{Probability of success after 200 steps with $n = 10$ and $N = 1024$ vertices. Shows the probability of success for $k$ adjacent marked vertices. \protect\subref{fig:success-probability-marked-neighbors-l-N-l-N-k-a} shows for $l = (n/N)$ and \protect\subref{fig:success-probability-marked-neighbors-l-N-l-N-k-b} for $l = (n/N)\cdot k$.}
\label{fig:success-probability-marked-neighbors-l-N-l-N-k}
\end{figure}

Figure \ref{fig:success-probability-marked-neighbors-l-N-l-N-k} shows the probability of success after two hundred steps. Figure \ref{fig:success-probability-marked-neighbors-l-N-l-N-k-a} shows the result for the value of $l = (n/N)$. The probability reaches its maximum when the number of vertices reaches $k = 4$ with a probability of success of $99.1\%$. Then the probability starts to decrease as $k$ increases. Figure \ref{fig:success-probability-marked-neighbors-l-N-l-N-k-b} shows the result for the value of $l = (n/N)\cdot k$. The probability reaches its maximum when the number of vertices reaches $k = 11$ with a success probability of $94.5\%$. Although the probability increases with a slower speed when $k = 5$, it already reaches $78.3\%$. This behavior is interesting for search spaces where the marked vertex density is high. Note the probability of the solid cyan curve. This behavior was found in work done by \citet{nahimovs2019adjacent} and was repeated here in our experiments. According to the authors, this is because the quantum walk has a stationary state.

\begin{figure}
\centerline{\includegraphics[width=0.7\textwidth]{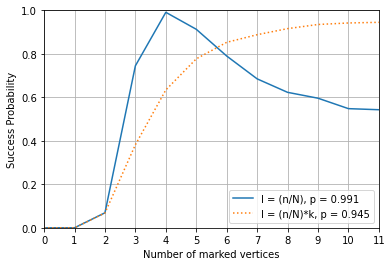}}
\caption{Maximum probability reached for each number of marked vertices in the neighborhood after one hundred steps with $n = 10$ and $N = 1024$ vertices. Evaluating the interference of the number of adjacent marked vertices in the value of $l$.} \label{fig:success-probability-l-n-and-l-n-k}
\end{figure}

Figure \ref{fig:success-probability-l-n-and-l-n-k} shows the comparison between what happens to the success probabilities in Figure \ref{fig:success-probability-marked-neighbors-l-N-l-N-k} when the number of $k$ increases. Note the dotted orange curve, the probability of success grows to its maximum value when the value of $l = (n/N)\cdot k$. The same does not happen when $l = (n/N)$.

We considered before that the marked vertices were neighbors. Now, let us analyze the possibility that in addition to having marked vertices in the neighborhood, there are also marked vertices that are not neighbors. We run ten experiments, and each one starts with two adjacent marked vertices $M = \{0,1\}$. In each experiment, a $i = \{1,2,3,\cdots\}$ non-adjacent vertex is randomly marked and the next marked neighbor, i.e., $M = \{0,1,2,...\}$. Therefore, in the tenth experiment, there will be eleven adjacent and ten non-adjacent vertices.

\begin{figure}
\subfloat[]{\includegraphics[width = 3.2in]{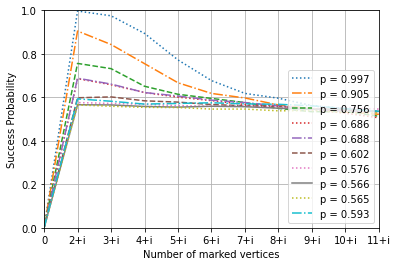}\label{fig:success-probability-adjacents-vertices-marked-not-marked-l-N-10-k-a}} 
\subfloat[]{\includegraphics[width = 3.2in]{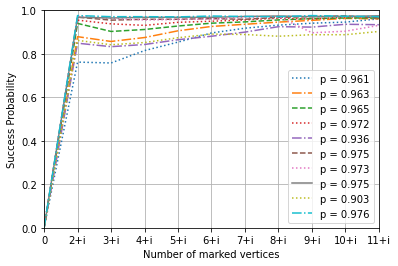}\label{fig:success-probability-adjacents-vertices-marked-not-marked-l-N-k-10-k-b}}
\caption{Maximum probability reached for each number of marked vertices after one hundred steps with $n = 10$ and $N = 1024$ vertices. \protect\subref{fig:success-probability-adjacents-vertices-marked-not-marked-l-N-10-k-a} shows the probability of success for $k$ adjacent and non-adjacent marked vertices for $l = (n/N)$. \protect\subref{fig:success-probability-adjacents-vertices-marked-not-marked-l-N-k-10-k-b} shows the probability of success for $k$ adjacent and non-adjacent marked vertices for $l = (n/N)\cdot k$.}
\label{fig:success-probability-adjacents-vertices-marked-not-marked}
\end{figure}

Figure \ref{fig:success-probability-adjacents-vertices-marked-not-marked} shows the behavior of probability amplitudes when for each set of adjacent vertices, a number of non-adjacent vertices are marked. Figure \ref{fig:success-probability-adjacents-vertices-marked-not-marked-l-N-10-k-a} shows that as new non-adjacent vertices are marked the probability is affected. Note that the behavior seen in the solid blue curve in Figure \ref{fig:success-probability-l-n-and-l-n-k} when there were no non-adjacent vertices is similar, i.e., as the density of the marked vertices increases, the probabilities decrease, even adding the vertices non-adjacent. The same can be seen in the case of the dotted orange curves in Figure \ref{fig:success-probability-l-n-and-l-n-k} and Figure \ref{fig:success-probability-adjacents-vertices-marked-not-marked-l-N-k-10-k-b}, i.e., when the density of the marked vertices increases, the probability also increases, this tells us that the value of $l = (n/N)\cdot k$ is optimal for high marked vertex densities. 

\begin{figure}
\subfloat[]{\includegraphics[width = 3.2in]{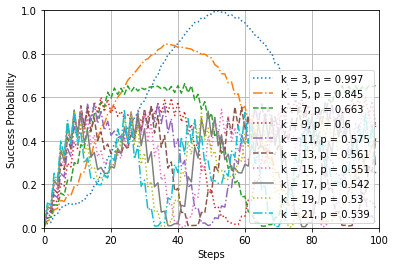}\label{fig:probability-of-success-forkadjacent-and-non-adjacent-marked-vertices-a}} 
\subfloat[]{\includegraphics[width = 3.2in]{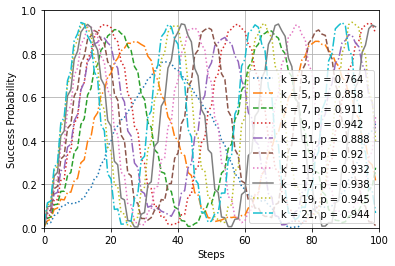}\label{fig:probability-of-success-forkadjacent-and-non-adjacent-marked-vertices-b}}
\caption{Probability of success after 100 steps with $n = 10$ and $N = 1024$ vertices. \protect\subref{fig:probability-of-success-forkadjacent-and-non-adjacent-marked-vertices-a}  shows the probability of success for $k$ adjacent and non-adjacent marked vertices for $l = (n/N)$. \protect\subref{fig:probability-of-success-forkadjacent-and-non-adjacent-marked-vertices-b} shows the probability of success for $k$ adjacent and non-adjacent marked vertices for $l = (n/N)\cdot k$.}
\label{fig:probability-of-success-forkadjacent-and-non-adjacent-marked-vertices}
\end{figure}

Figure \ref{fig:probability-of-success-forkadjacent-and-non-adjacent-marked-vertices} shows the probability of success for the search of marked adjacent and non-adjacent vertices in the search space. We performed an experiment, where, at every hundred steps, an adjacent vertex and a non-adjacent vertex were marked, i.e., for each $M$ set of adjacent vertices a vertex $i \notin M$ was marked randomly, then, $M' = \{0,1,i_{0}\},\{0,1,i_{0},2,i_{1}\},\cdots,\{0,1,i_{0},2,i_{1},4, i_{2},\cdots,512,i_{10}\}$. Figure \ref{fig:probability-of-success-forkadjacent-and-non-adjacent-marked-vertices-a} shows the probability of success for $l = (n/N)$ and Figure \ref{fig:probability-of-success-forkadjacent-and-non-adjacent-marked-vertices-b} shows the probability of success for $l = (n/N) \cdot k$. Note that the probability of success above $90\%$ is achieved in a smaller number of steps.

\section{Conclusions}
\label{sec:conclusions}

Many efforts are applied in order to improve the performance of quantum search algorithms. Quantum walks are the main tool for building these algorithms. We initially analyzed the quantum walk in the hypercube applying Grover's search and came to the conclusion that neighbor vertices affect the search performance, an observation that has been corroborated by other authors. We found that the walk could not improve its results even for a number of marked vertices equal to one. Many authors have developed works for adding self-loops in various types of graphs and grids of different dimensions. In this sense, we decided to investigate how to improve the quantum search in the hypercube using self-loops. Previous works defined the optimal self-loop value as $l = (d/N)$ for one marked vertex to the quantum walk on the hypercube. After performing experiments we saw that this value of $l$ was not optimal for multiple marked vertices. We arrive at a value of $l = (n/N)\cdot k$ for an arbitrary number of vertices. This value is also used when searching in two-dimensional grids. Another aspect of the quantum walk in the hypercube is whether the marked vertex is adjacent or not, this interferes with the search performance. We then analyzed whether the value of $l = (n/N)$ and $l = (n/N)\cdot k$ had any positive effect when applied to the hypercube vertices. The results show that the value of $l = (n/N)$ is not optimal for the quantum walk in the hypercube with multiple marked vertices adjacent or not. It also shows that for a search space where there are marked adjacent vertices, just one non-adjacent marked vertex is sufficient for the value of $l = (n/N)\cdot k$ to be better. According to the results presented here, there is a greater than $90\%$ probability that the measurement will collapse in one of the solutions. Recent works have used the quantum walks to search weights and train artificial neural networks \citep{de2019quantum,desouza2021classical}. The quantum walk in the hypercube has an interesting behavior, the amplification of neighbors vertices. In future work, we intend to use this quantum walk to find a set of weights to initialize and train classical artificial neural networks. We also intend to analyze the quantum walk in the hypercube with multiple weighted self-loops.

\section*{Acknowledgments}
\label{sec:acknowledgments}
Acknowledgments to the Science and Technology Support Foundation of Pernambuco (FACEPE) Brazil, Brazilian National Council for Scientific and Technological Development (CNPq), and Coordena\c{c}\~{a}o de Aperfei\c{c}oamento de Pessoal de N\'{i}vel Superior - Brasil (CAPES) - Finance Code 001 by their financial support to the development of this research.

\bibliographystyle{unsrtnat}






\end{document}